\documentclass[twocolumn,showpacs,preprintnumbers,amsmath,amssymb,prl]{revtex4}

\usepackage{graphicx}
\usepackage{dcolumn}
\usepackage{bm}

\begin{document}

\preprint{APS/123-QED}

\title{Growth of (Ge,Mn) nanocolumns on GaAs(100): the role of morphology and co-doping on magnetotransport}

\author{M. Jamet$^{1}$, I.-S. Yu$^{1}$, T. Devillers$^{1}$, A. Barski$^{1}$, P. Bayle-Guillemaud$^{1}$, C. Beigne$^{1}$, J. Rothman$^{2}$, V. Baltz$^{3}$, J. Cibert$^{4}$}
\affiliation{$^{1}$INAC/SP2M, CEA-UJF, Grenoble, France\\
$^{2}$LETI/DOPT/LIR, CEA, Grenoble, France\\
$^{3}$INAC/Spintec, CEA-CNRS-UJF, Grenoble, France\\
$^{4}$Institut N\'eel, CNRS-UJF, Grenoble, France}%

\date{\today}

\begin{abstract}
Changing the morphology of the growing surface and the nature of
residual impurities in (Ge,Mn) layers - by using different
substrates - dramatically changes the morphology of the
ferromagnetic Mn-rich inclusions and the magnetotransport
properties. We obtained p-type layers with nanocolumns, either
parallel or entangled, and n-type layers with spherical clusters.
Holes exhibit an anomalous Hall effect, and electrons exhibit a
tunneling magnetoresistance, both with a clear dependence on the
magnetization of the Mn-rich inclusions; holes exhibit orbital MR,
and electrons show only the normal Hall effect, and an additional
component of magnetoresistance due to weak localization, all three
being independent of the magnetic state of the Mn rich inclusions.
Identified mechanisms point to the position of the Fermi level of
the Mn-rich material with respect to the valence band of germanium
as a crucial parameter in such hybrid layers.
\end{abstract}

\maketitle

Research on ferromagnetic semiconductors triggered enormous activity
due to their potential use in spintronics
\cite{MacDonald2005,Dietl2002}. Up to now, efforts have mainly
focused on diluted magnetic semiconductors (DMS) in which magnetic
atoms randomly substitute the host matrix atoms \cite{Ohno1998}.
Their magnetic properties can be manipulated by electric fields
making them suitable materials for spintronic applications provided
that they can be made ferromagnetic above room temperature. However
DMS based on II-VI and III-V semiconductors still exhibit very low
values of Curie temperature $T_C$.

Many groups have reported $T_C$ values well above room temperature,
along with remarkable magneto-transport and magneto-optical properties,
in semiconductors doped with magnetic transition metals (TM). It is
now admitted that these properties may be attributed to TM-rich
areas resulting from spinodal decomposition \cite{Dietl2008}. Such
features have been theoretically predicted \cite{Sato2005} and
reported in (Ge,Mn)
 \cite{Jamet2006,Bougeard2006,Li2007,Devillers2007}, and in Cr and
Fe-doped GaN \cite{Gu2005,Bonanni2008} or ZnTe \cite{Kuroda2007}. In
this field of intense materials research, goals are now: (i)
controlling spinodal decomposition to reproducibly stabilize
high-T$_{C}$ TM-rich areas and tailor desirable magnetic properties,
and (ii) enhancing the coupling with carriers to give rise to strong
magneto-resistance (MR) or anomalous Hall effect (AHE).

In this paper, we demonstrate the fine control of spinodal
decomposition in (Ge,Mn) films grown on GaAs(001) substrates. We
focus on (Ge,Mn) because it is compatible with mainstream silicon
technology, and spinodal decomposition leads to high $T_C$ values in
layers grown on Ge substrate. Growing (Ge,Mn) films on GaAs(001)
semi-insulating ($\rho > 10^7~\Omega$cm) substrates makes in-plane
transport measurements easier, and constitutes a first step towards
spin injection from (Ge,Mn) into a GaAs-based spin-LED
\cite{Fiederling1999}. Using different surface preparations, we
clearly identify the role of surface morphology and the role of
impurity diffusion from the substrate (either Ga or As atoms), on
the nanocolumns growth, on one hand, and on the electrical
properties, on the other hand. We thus address the major issue of
the influence of co-doping (either n-type or p-type) on spinodal
decomposition in group IV magnetic semiconductors, demonstrating a
major influence on the shape of the Mn-rich precipitates. We also
provide new hints to control and optimize magneto-transport
propertie of the (Ge,Mn) films. We show that AHE and MR are not
optimized simultaneously, and we propose a general picture based on
the electrical doping of the matrix and on the position of Fermi
level in the precipitates with respect to the valence band of Ge.

(Ge,Mn) films were grown by low temperature Molecular Beam Epitaxy
(MBE), using growth conditions as described in
Ref.~\cite{Jamet2006}, with the substrate temperature
$T_{g}=100^{\circ}$C and deposition rate $\sim$ 0.2~\AA.s$^{-1}$. We
have used two different methods to prepare the GaAs surface. In the
first one, the native oxide was thermally desorbed from an
$\it{epiready}$ substrate, by raising the substrate temperature up
to almost 600$^{\circ}$C. The (Ge,Mn) layers was grown directly on
the resulting Ga-rich GaAs surface, which was rough as observed by
RHEED. Such samples will be labelled Ga-(Ge,Mn). In the second case,
As-(Ge,Mn) samples, a thin undoped GaAs buffer layer was grown first
in a separate III-V system, protected with an amorphous As capping,
and transferred in air to the IV-IV MBE machine. Desorbing the As
capping layer at 200$^{\circ}$C results in a very flat, (2$\times$4)
reconstructed, As-rich surface. Samples grown on Ge substrates,
labelled Ge-(Ge,Mn), constitute our reference samples.

Magnetization was measured using a Superconducting QUantum
Interference Device (SQUID). Magnetotransport properties
(magnetoresistance and Hall effect) were investigated using Hall
bars defined by optical lithography, aligned along a $<110>$
direction, of width 20 $\mu$m, with voltage probes separated by 140
$\mu$m.

\begin{figure}[h!]
\begin{center}
\includegraphics[width=0.42\textwidth]{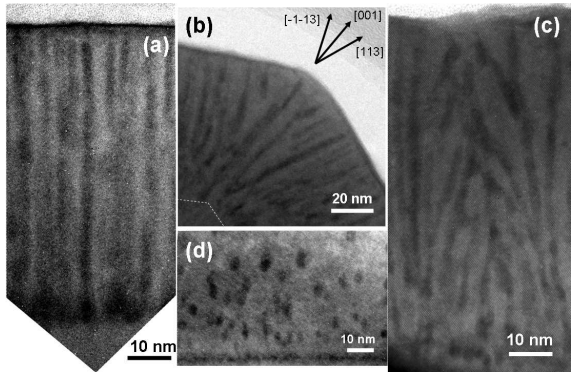}
\end{center}
\caption{TEM cross section of 80~nm thick layers : (a)
Ge$_{0.94}$Mn$_{0.06}$ grown at 100$^{\circ}$C after depositing a
40~nm thick Ge buffer on GaAs(001) and (b) at 100$^{\circ}$C on the
facetted Ge surface. (c) Ga-Ge$_{0.9}$Mn$_{0.1}$ and (d)
As-Ge$_{0.98}$Mn$_{0.02}$ films grown at 100$^{\circ}$C.}
\label{fig1}
\end{figure}

A typical morphology is that of long Mn-rich nanocolumns, growing
normal to the substrate surface. On a Ge substrate \cite{Jamet2006}
or on a Ge buffer layer grown on GaAs(001) (Fig.~1a), these nanocolumns are well
aligned along the [001] growth direction. On a Ge buffer layer grown
on a Ge(001) substrate with \{113\} facets obtained by anisotropic
chemical etching in an H$_2$O$_2$ aqueous solution, Fig.~1b, they
grow perpendicular to the facets. Finally, in Ga-(Ge,Mn) films,
they are bent according to the initial surface roughness (Fig.~1c),
and this results in a highly disordered pattern. This general
picture fully agrees with 2D spinodal decomposition, driven by
surface diffusion and aggregation of Mn atoms, with nucleation of
Mn-rich areas taking place during the first stage of the growth
\cite{Sato2005}. As a consequence of this mechanism, the columns are
always perpendicular to the growing surface.

\begin{figure}[h!]
\begin{center}
$\begin{array}{cc}
\includegraphics[width=0.24\textwidth]{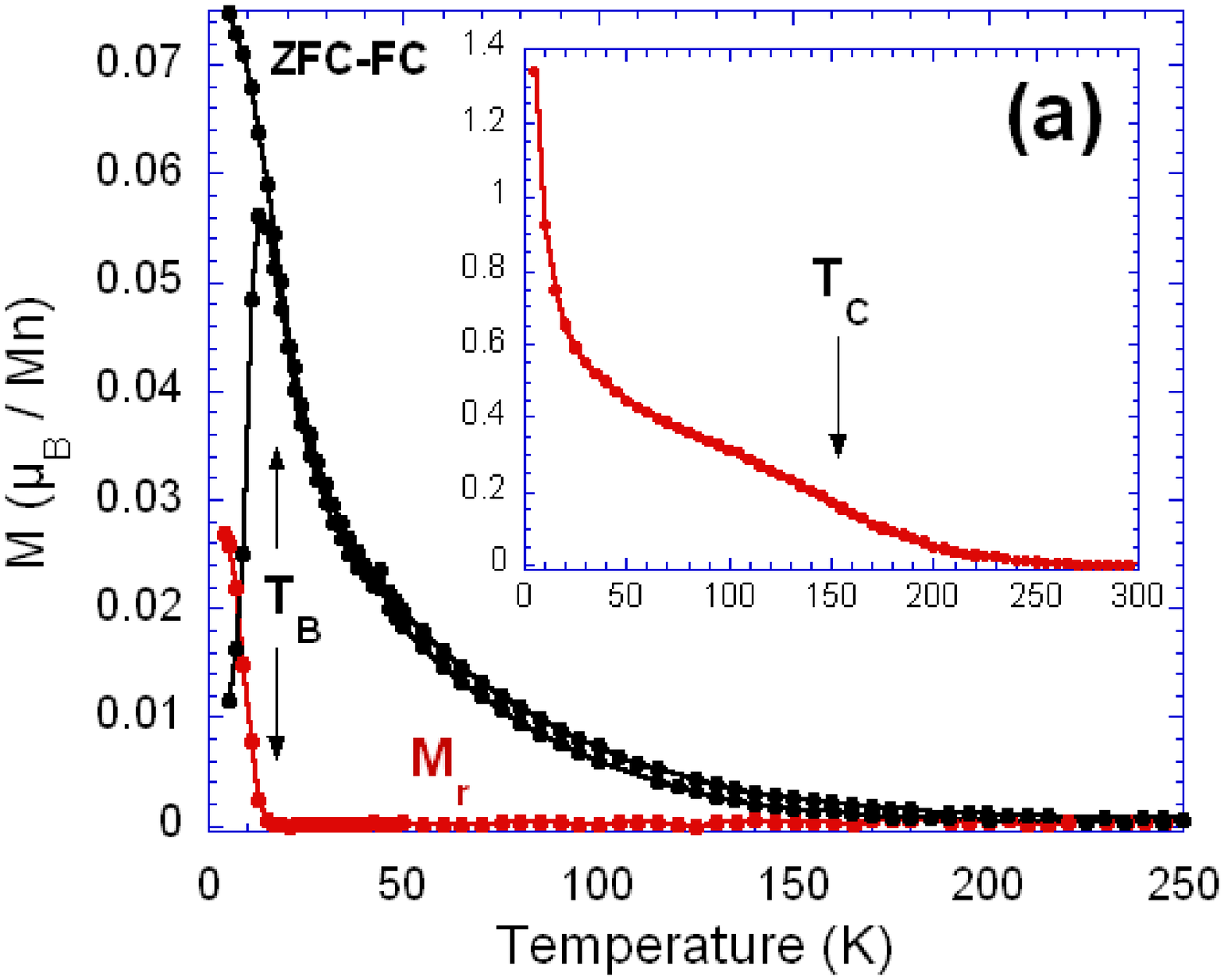}  &
\includegraphics[width=0.24\textwidth]{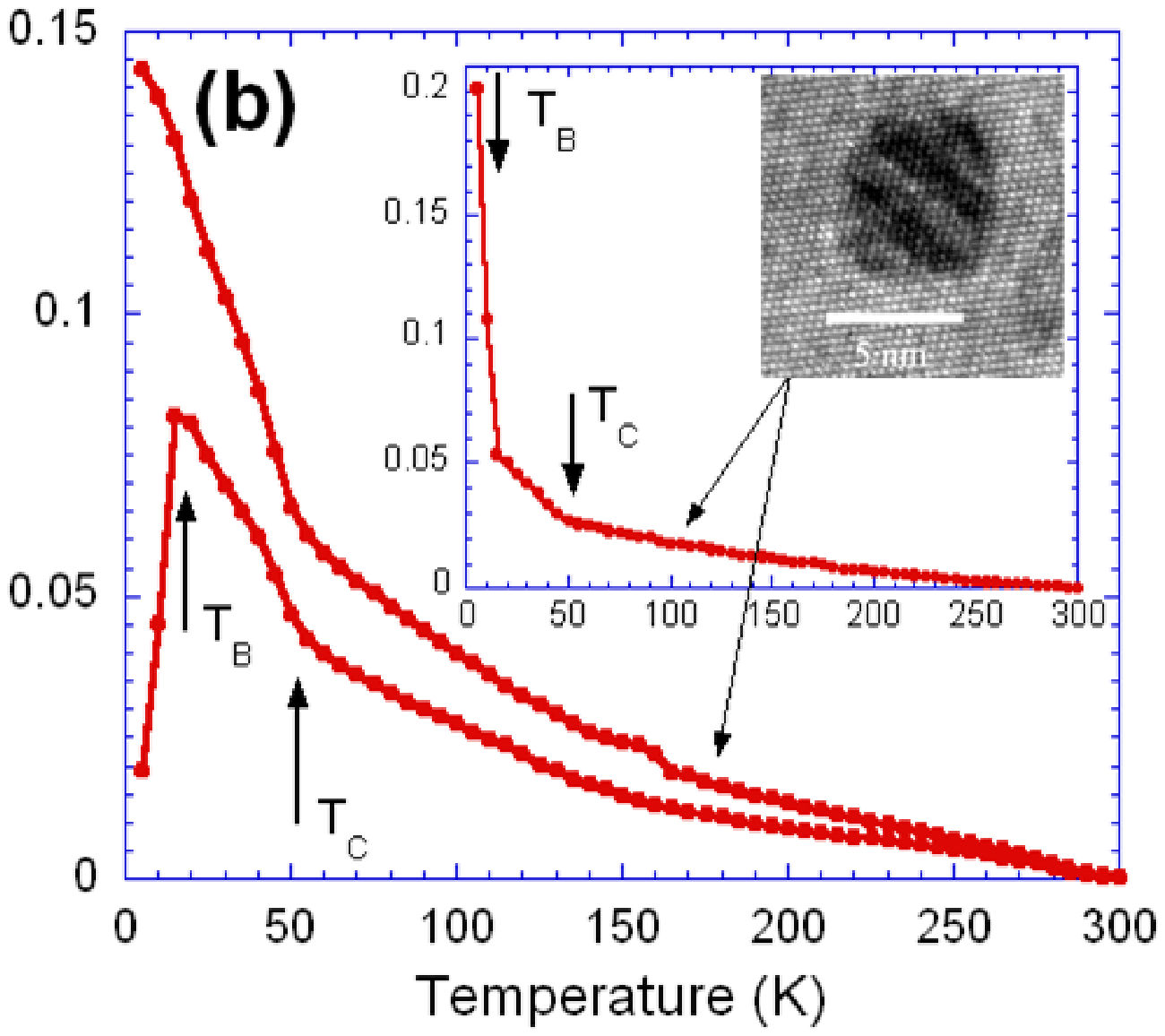}
\end{array}$
\end{center}
\caption{Temperature dependence of magnetization in (a)
Ga-Ge$_{0.9}$Mn$_{0.1}$ and (b) As-Ge$_{0.94}$Mn$_{0.06}$: ZFC-FC
curves at 0.015~T and magnetic remanence $M_r$ after maximum field
cooling at 5~T. Inset: in (a), saturation magnetization at 2~T, and
in (b), TEM image of a Ge$_{3}$Mn$_{5}$ cluster.} \label{fig2}
\end{figure}

By contrast, in Fig.~1d, the As-(Ge,Mn) layers feature randomly
distributed Mn-rich precipitates. In addition, a few
Ge$_{3}$Mn$_{5}$ clusters already start to form, as evidenced by
their typical Moir\'e contrast (see inset of Fig.~2b). The same
random distribution of nanoclusters is observed when increasing the
nominal Mn content from 2\% to 6\% and 10\% at
$T_{g}$=100$^{\circ}$C, but for an increase of the average
precipitate density and length along the growth direction. This
suggests that the decomposition is of 3D character and mostly driven
by nucleation.

Secondary Ion Mass Spectroscopy (SIMS, not shown) performed on the
As$-$(Ge,Mn) samples evidences an As-rich topmost layer, extending
over 3$\pm$1~nm below the sample surface and containing up to 6$\times
10^{19}$ As cm$^{-3}$, \emph{i.e.}, an integrated amount of almost 1
ML As. This is a consequence of the segregation of As atoms,
initially present at the GaAs surface, during the growth of Ge, with
a well-known surfactant effect \cite{Leycuras1995}. Accordingly, as
described below, the As-(Ge,Mn) films appear as n-doped: As atoms
are shallow donors in Ge, and in this topmost layer they compensate
p-type doping by substitutional Mn.

The presence of As near the surface of the growing layer offers a
possible explanation for this change of character of the spinodal
decomposition, from 2D to 3D. According to Ref.~\cite{Zhu2004}, Mn
atoms are incorporated into germanium in a subsurface interstitial
position, and further diffuse within the growth plane: this offers a
mechanism for 2D spinodal decomposition \cite{Sato2005}. Codoping
with As changes the charge state of Mn atoms, thus reducing Coulomb
repulsion and enhancing the effect of attractive Mn-Mn pair
interaction, making the nucleation of Mn-rich precipitates easier
\cite{Dietl2008}. In addition to that mechanism, the presence of
donors like As is expected to displace the equilibrium between
interstitial Mn (another donor) and substitutional Mn (an acceptor),
enhancing the amount of substitutional Mn (which form nucleation
centers for further Mn aggregation \cite{Zhu2008}), and reducing the
amount of interstitial Mn (thus decreasing the incorporation into
already existing clusters). These different mechanisms induced by
the presence of As conspire to favor a growth process dominated by
nucleation, contributing to make the spinodal decomposition 3D.

A complete study of the magnetic properties will be published
elsewhere \cite{comment2}. All samples exhibit two [case of
Ga-(Ge,Mn), Fig.~2a] or three [case of As-(Ge,Mn), Fig.~2b] magnetic
phases, as evidenced from the temperature dependence of the
saturation magnetization $M_s$ (inset of Fig.~2) and the remanent
magnetization $M_r$, and the ZFC-FC curves. They exhibit: (i) a
strong paramagnetic signal with a $1/T$ temperature dependence at
low temperature, attributed to Mn atoms diluted in the Ge matrix,
and well fitted using a 3/2-Brillouin function
\cite{Schulthess2001}; (ii) a contribution attributed to the
superparamagnetic Mn-rich nanocolumns or precipitates, with finite
$T_C$ and blocking temperature $T_B$; (iii) in As-(Ge,Mn) only, a
contribution from Ge$_{3}$Mn$_{5}$ clusters with a broad range of
blocking temperatures. A detailed analysis \cite{comment2} of the
Ga-(Ge,Mn) sample in Fig.~2a shows that 40$\pm$6\% of the magnetic
moments are in nanocolumns, with $T_C\approx150$~K and $T_B=15
\pm$5~K, 1.0$\pm$0.1~$\mu_{B}$/Mn, and an average magnetic moment of
a nanocolumn 520$\pm$50~$\mu_{B}$. For the As-(Ge,Mn) sample in
Fig.~2b, we found that 52$\pm$3\% of the magnetic moments are in the
matrix, and 22$\pm$2\% in the Mn-rich precipitates with
$T_{C}\approx$~50~K, $T_{B}$=15$\pm$5~K, 1.2$\pm$0.2~$\mu_{B}$/Mn,
and $\approx$100$\pm$20~$\mu_{B}$ per precipitate.

\begin{figure}[h!]
\begin{center}
$\begin{array}{cc}
\includegraphics[width=0.23\textwidth]{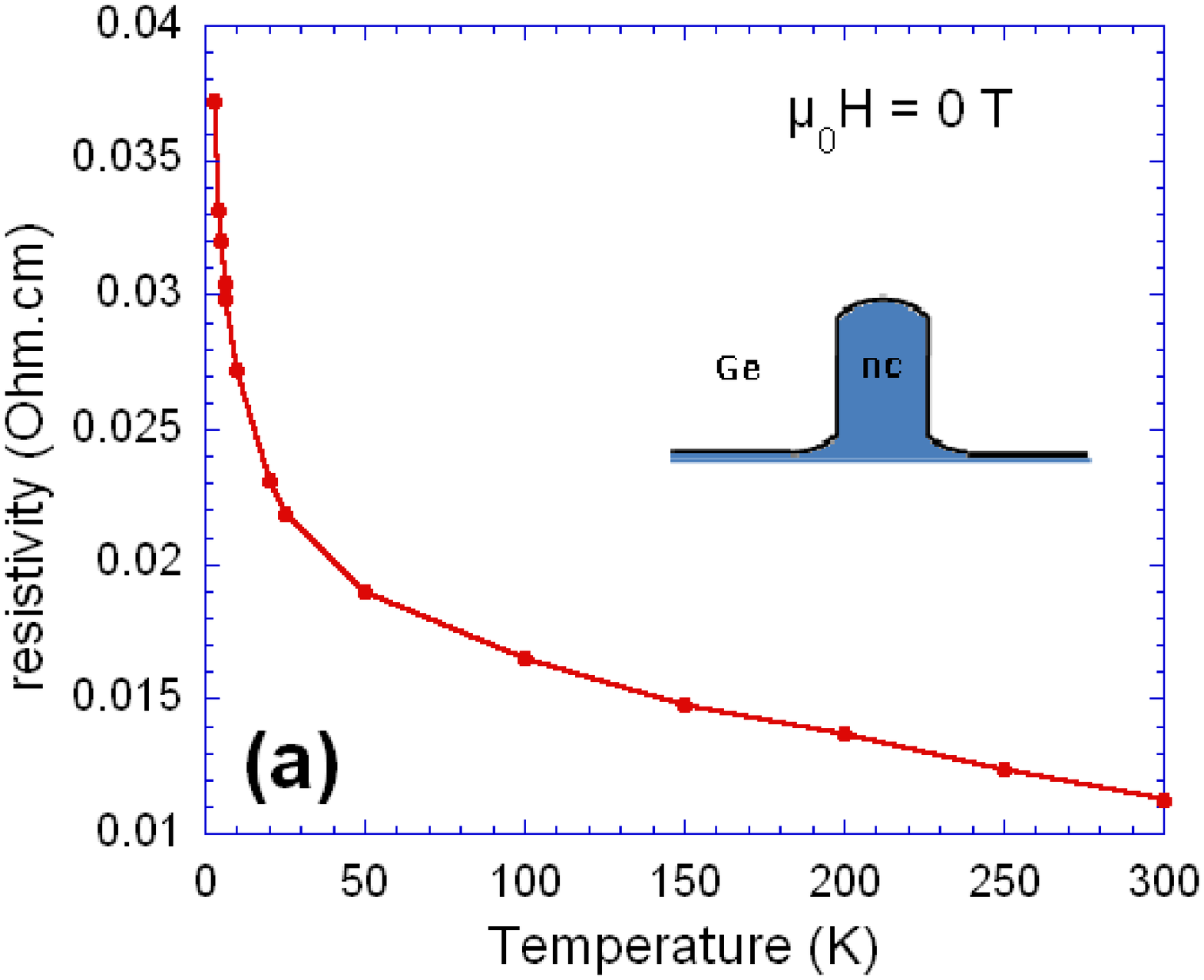}  &
\includegraphics[width=0.23\textwidth]{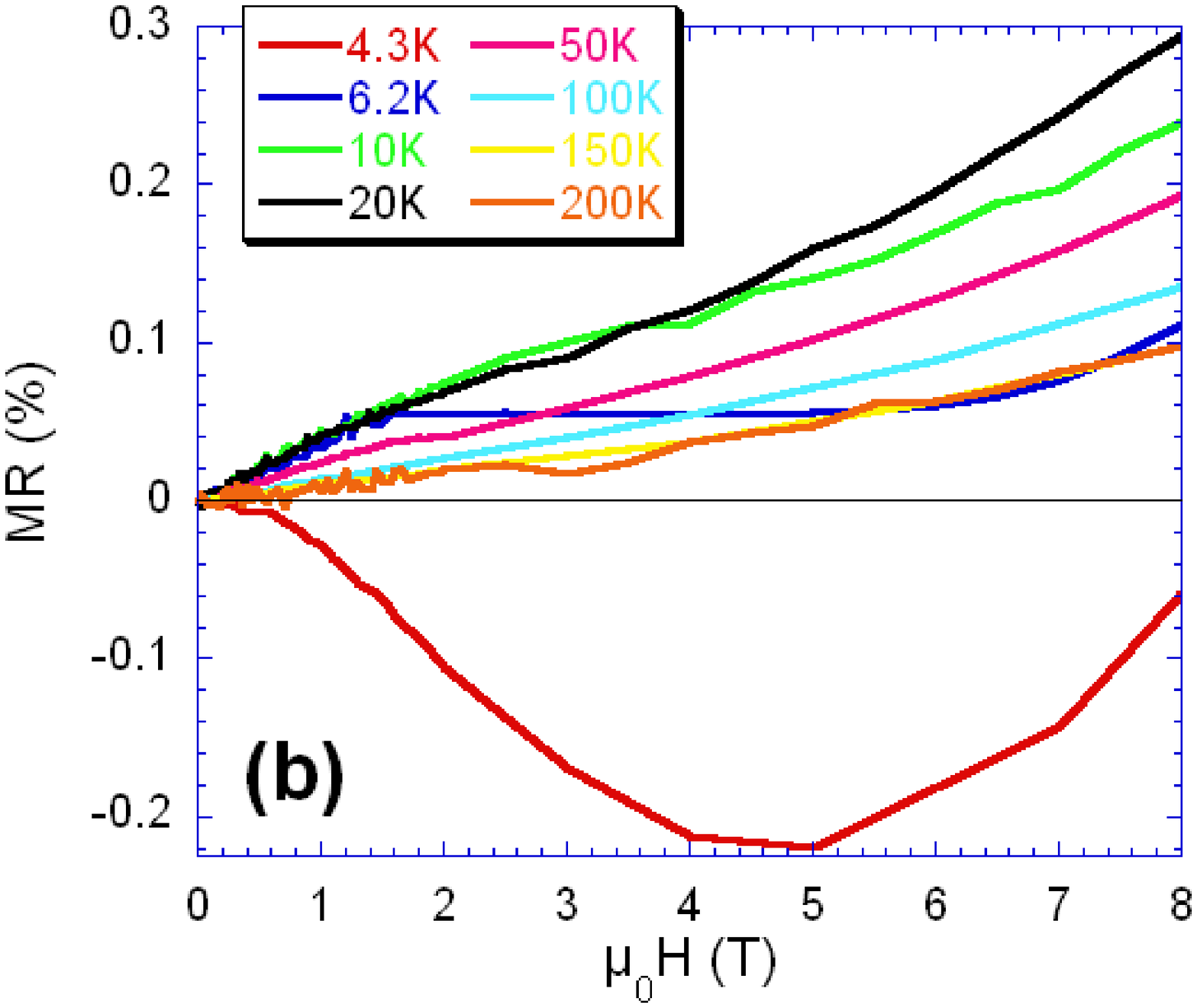} \\
\includegraphics[width=0.23\textwidth]{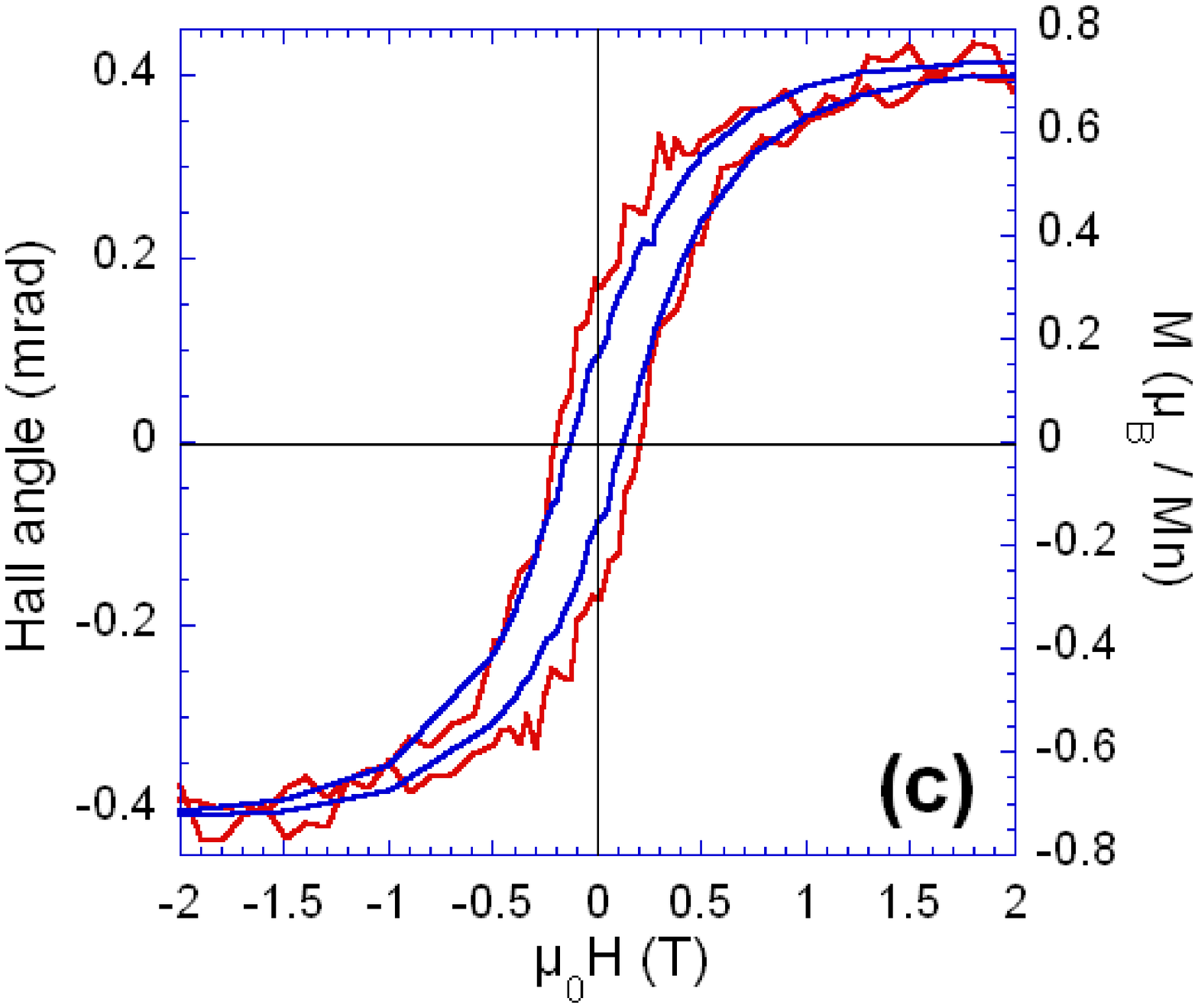}  &
\includegraphics[width=0.23\textwidth]{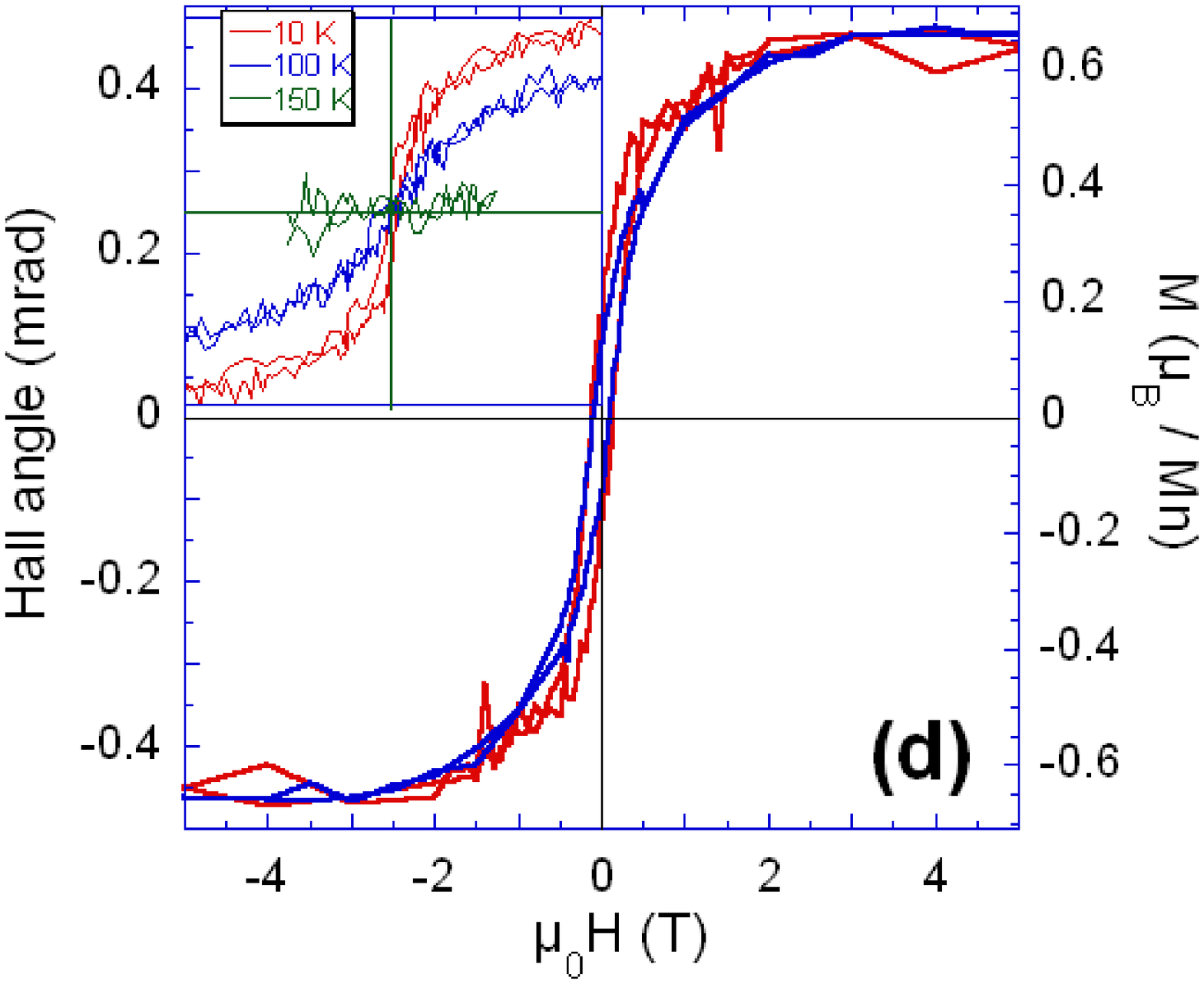}
\end{array}$
\end{center}
\caption{Magnetotransport in Ga-Ge$_{0.9}$Mn$_{0.1}$, magnetic field
applied normal to the plane: temperature dependence of the
zero-field resistivity (a), MR up to 8~T at different temperatures
(b), and AHE (red) compared to magnetization (blue) at 2-3 K (c) and
4-5 K (d). Inset: AHE at 10, 100 and 150~K.} \label{fig3}
\end{figure}

From the high field slope of the Hall effect,
Ga-Ge$_{0.9}$Mn$_{0.1}$ films are p-type, with a hole density up to
3$\times$10$^{19}$~cm$^{-3}$ at 300~K, comparable to what we gave as
a lower bound (due to a strong AHE) of the hole density in
Ge-(Ge,Mn). The resistivity (Fig.~3a) is of the insulating type (it
was metallic in Ge-(Ge,Mn)). Mn was reported as a double acceptor in
germanium, with acceptor levels 160~meV and 370~meV above the
valence band edge \cite{Woodbury1955} respectively. For such a deep
acceptor, the Mott critical density is expected to be well in the
10$^{20}$~cm$^{-3}$ range, and for a measured density one order of
magnitude lower we should observe a strongly activated conductivity
\cite{Achatz2008}: we actually observe (Fig.~3a) a weak
temperature dependence. We conclude that the Hall effect is not due
solely to the holes from the Mn acceptors in the germanium matrix.
This is possible if the Fermi level of the Mn-rich material forming
the precipitates lies below the top of the valence band in
germanium, so that no Schottky barrier, and even an accumulation
layer, is formed around each precipitate. That induces the build-up
of an electric field pattern around each nanocolumn, which drags
holes towards the nanocolumn.

Then the magnetotransport properties can be understood as follows:
(i) as the nanocolumns configuration is well below the percolation
threshhold, holes have to propagate through the germanium matrix;
that makes the basis of the conductivity; (ii) however, the electric
field pattern drags the holes through the nanocolumns, where the
conductivity is higher; applying a magnetic field suppresses this
effect, creating the geometrically enhanced orbital MR, or
Extraordinary MR (EMR)\cite{Solin2000} which we observed to be strong
in Ge-(Ge,Mn) \cite{Jamet2006}; (iii) finally, the absence of
Schottky barrier enhances the interaction of holes with Mn atoms
in the nanocolumns, thus allowing a spin polarization and a strong
AHE to appear \cite{Jamet2006}.

In the Ga-(Ge,Mn) samples, we still observe the EMR (Fig.~3b), with
the same temperature dependence as in Ref.~\cite{Jamet2006}, but
much weaker (although much higher than classical Lorentz MR). This is readily explained by considering the dependence
of EMR on the carrier mobility (it scales as $\mu^{2}$): as seen in
Fig.~3a, the zero-field resistivity $\rho_{0}$ in Fig.~3a is of the
order of 10$^{-2}$ $\Omega$~cm, so that the mobility is lower than
in Ge-(Ge,Mn) by more than one order of magnitude, possibly due to
the higher disorder and induced defects. Finally, at very low
temperature, negative MR is observed, which may be due to spin
disorder scattering \cite{Matsukura1998} on  Mn atoms diluted in the
Ge matrix (as typical in metallic DMS), or to GMR on the Mn-rich
nanocolumns. In both cases, this effect is expected to be very weak;
in addition the spin diffusion length of holes is very short due to
spin-orbit coupling, making GMR on Mn-rich nanocolumns unlikely.

AHE in Ga-(Ge,Mn) nicely matches the magnetization of the
nanocolumns (Fig.~3c-d). Again, the effect is weaker than in
Ge-(Ge,Mn), by almost two orders of magnitude. This is expected from
the lower mobility: it was pointed out in Ref.~\cite{Park2002} that
scattering on impurities such as Ga atoms (SIMS measurements have indeed shown that Ga out-diffused from the GaAs substrate) partly suppresses the effect of skew scattering.

\begin{figure}[h!]
\begin{center}
$\begin{array}{cc}
\includegraphics[width=0.23\textwidth]{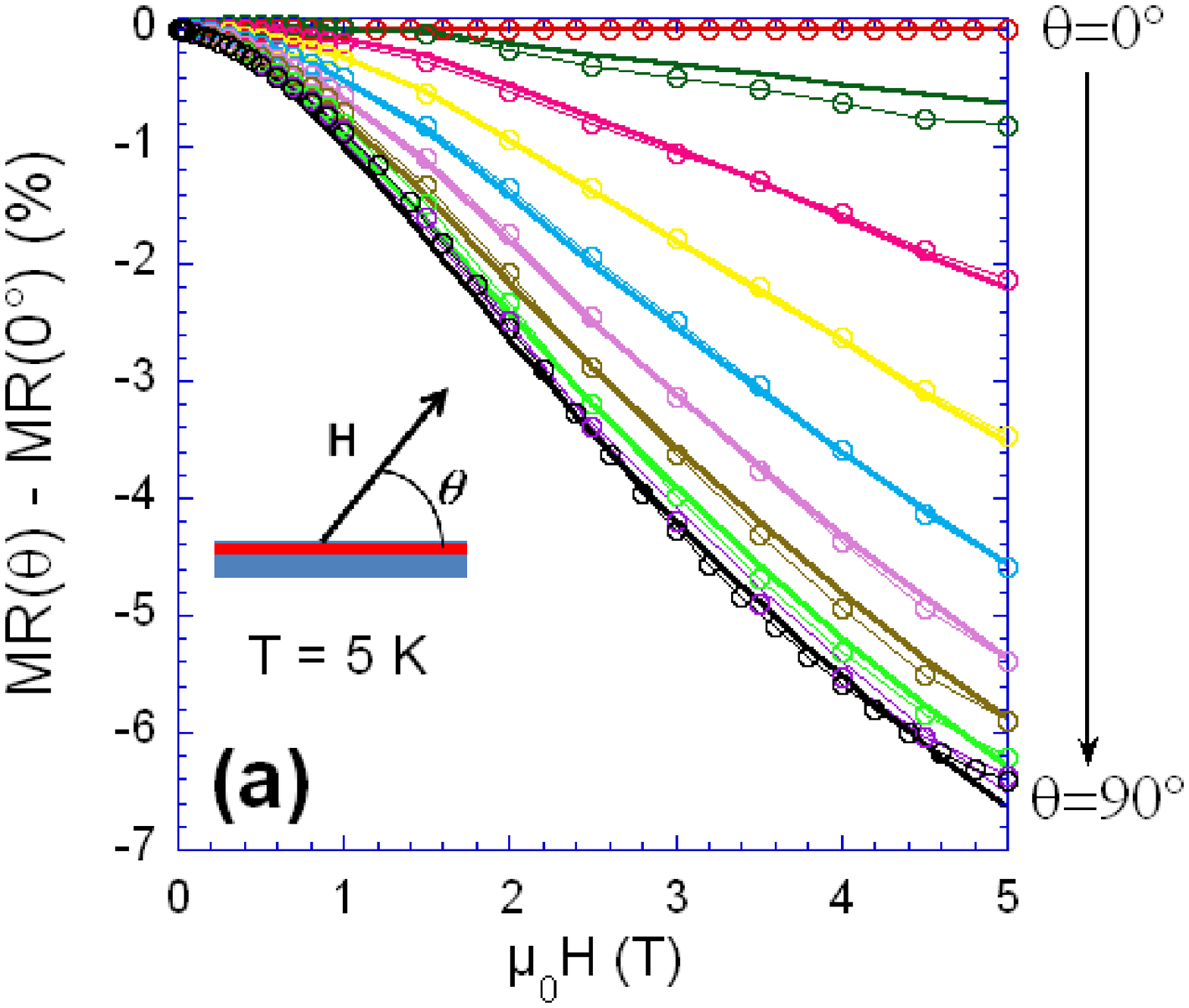}  &
\includegraphics[width=0.23\textwidth]{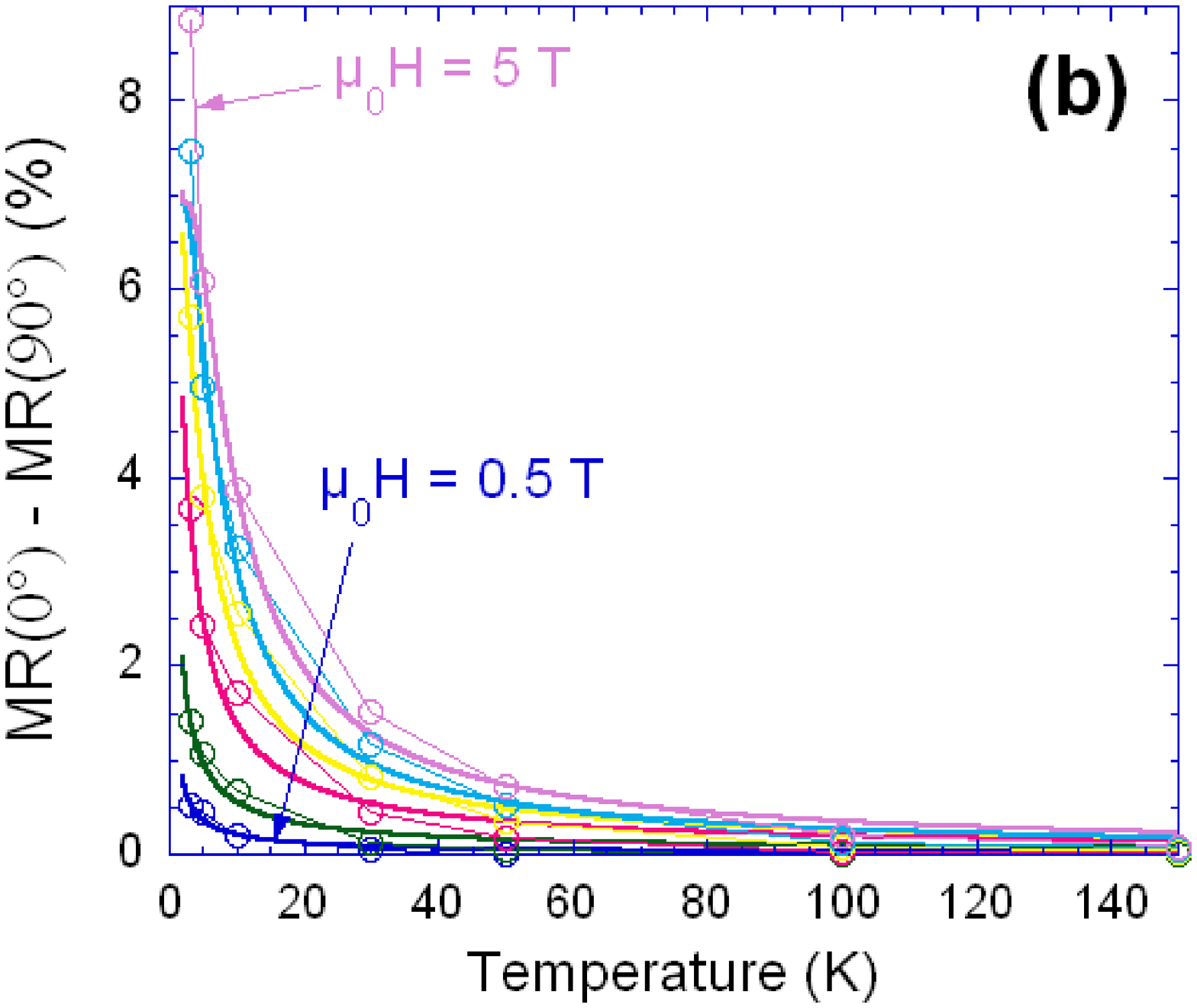} \\
\includegraphics[width=0.23\textwidth]{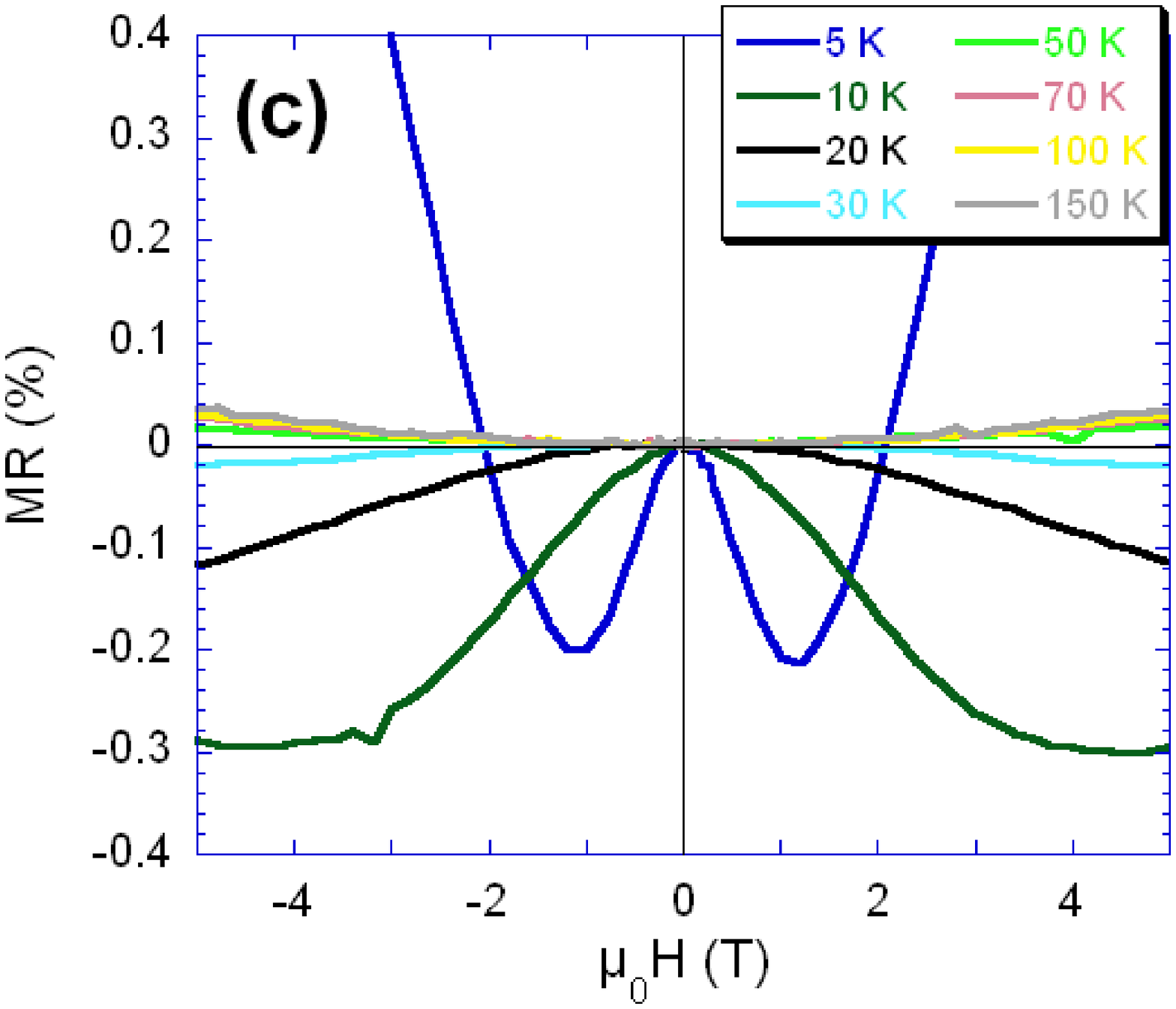}  &
\includegraphics[width=0.23\textwidth]{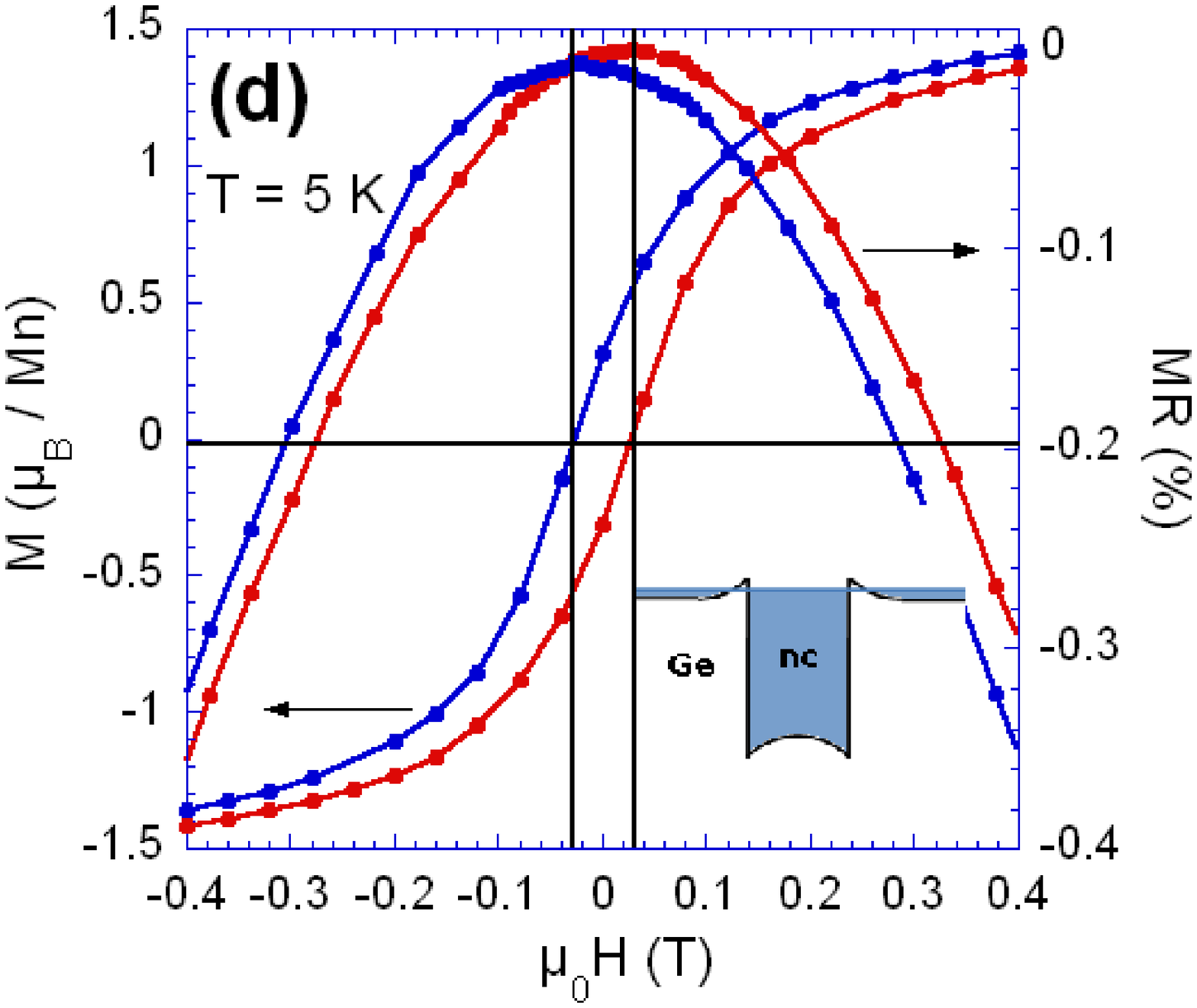}
\end{array}$
\end{center}
\caption{Magnetotransport in As-Ge$_{0.94}$Mn$_{0.06}$. (a) field
dependance and (b) temperature dependence of the MR anisotropy, for
different orientations of the field; symbols are experimental data,
solid line calculated. (c,d) MR and magnetization as a function of
the field applied in-plane.} \label{fig4}
\end{figure}

As-(Ge,Mn) films exhibit metallic n-type conductivity. This is
clearly due to the presence of As donors in the topmost layer. Hence
magnetotransport essentially measures the properties of this 3~nm
thick layer. We observe no AHE, as expected since spin-orbit
scattering is small for electrons in the conduction band of
germanium, and also because the same assumptions as above (Fermi
level of precipitates lying below the top of the valence band)
creates a high Schottky barrier for electrons.

MR is highly anisotropic (Fig.~4a): we show now that this is due to
2D weak localization in the As-doped layer, which vanishes when the
field is applied in-plane ($\theta=0$). The isotropic part
(Fig.~4c-d) will be analyzed later on.

The MR of non-interacting electrons in the 2D weak localization
regime is $\Delta \rho/\rho \approx -\Delta \sigma/\sigma =
-A.f_{2}[4e \mu_0 H sin(\theta) L_\phi /\hbar]$
\cite{Altshuler1985}, where $A=e^{2}/2\pi^{2}\hbar\sigma_{2D}(0)$,
$L_\phi$ is the phase relaxation length, and the function $f_{2}(x)$
is defined in \cite{Altshuler1985}. Here we neglect the effect of
spin-orbit coupling and the anisotropy of $L_\phi$ \cite{comment2}.
Fits in Fig.~4a were obtained with only two adjustable parameters:
$A=4\%$, close to the value $A=(6\pm2)\%$ calculated using the
experimental value of the 2D conductivity at $H=0$, and
$L_{\phi}=11.5$~nm, large enough with respect to the thickness of
the conducting layer to justify the use of the 2D regime of weak
localization. Moreover, fits of the temperature dependence of the
anisotropic MR (Fig.~4b) were obtained by simply writing
$L_\phi=\sqrt{D\tau_{\phi}}$, using the temperature dependence of
the diffusion coefficient $D$ for an n-doped degenerate
semiconductor \cite{Sze1981}, and a temperature dependence of the
phase relaxation time $\tau_{\phi}\propto T^{-\alpha}$ with
$\alpha\approx$1.7, similar to that obtained in
\cite{Polyanskaya1981} for Ge:Sb ($\alpha$=1.5) and in
\cite{Uren1981} for Si-MOSFETs ($\alpha$=1.6), and currently
attributed to both electron-electron and electron-phonon collisions.

Turning back to the isotropic MR, Fig.~4c-d, it contains a negative
contribution, which features two maxima at the coercive field of
Mn-rich precipitates, and vanishes above 50~K as does their
magnetization. Hence, we tentatively ascribe it to tunneling MR
(TMR) through the precipitates and the Schottky barriers formed
around them. By analogy with spin injection from a ferromagnetic
metal to a semiconductor \cite{Fert2001}, efficient spin injection
from the precipitate to the matrix requires an interface resistance
provided by the Schottky barrier. This barrier must be high enough
to prevent full spin relaxation inside the precipitate but
reasonably transparent to allow tunnel MR to occur.

To summarize, we have shown that surface morphology and co-doping
have major influence on spinodal decomposition in (Ge,Mn) films
grown on GaAs(001) substrates. For films grown on Ga-rich rough
surfaces, we recovered 2D spinodal decomposition with bent
nanocolumns. Electrical properties are similar to what we obtained
on Ge(001) substrates except that the presence of defects in the
films leads to weaker positive MR and AHE. For films grown on
As-rich flat surfaces, 3D spinodal decomposition is observed due to
As co-doping and magnetotransport is dominated by TMR and weak
localization, AHE is negligible. These results are consistent with
the assumption that the Fermi level of Mn-rich precipitates lies in
the valence band of the Ge matrix.

We thank A. Arnoult (LAAS, Toulouse) for providing As-capped GaAs
substrates, T.~Dietl and J.~Pernot for fruitful discussions.
This work was granted by the Agence Nationale pour la Recherche
(project GeMO) and the Nanoscience Foundation in Grenoble (project
IMAGE).


\begin{thebibliography}{}
\expandafter\ifx\csname natexlab\endcsname\relax\def\natexlab#1{#1}\fi
\expandafter\ifx\csname bibnamefont\endcsname\relax
  \def\bibnamefont#1{#1}\fi
\expandafter\ifx\csname bibfnamefont\endcsname\relax
  \def\bibfnamefont#1{#1}\fi
\expandafter\ifx\csname citenamefont\endcsname\relax
  \def\citenamefont#1{#1}\fi
\expandafter\ifx\csname url\endcsname\relax
  \def\url#1{\texttt{#1}}\fi
\expandafter\ifx\csname urlprefix\endcsname\relax\def\urlprefix{URL }\fi
\providecommand{\bibinfo}[2]{#2}
\providecommand{\eprint}[2][]{\url{#2}}

\bibitem{MacDonald2005}
\bibinfo{author}{\bibfnamefont{A. H.}~\bibnamefont{MacDonald \textit{et al.} }},
  \bibinfo{journal}{Nature Mater.} \textbf{\bibinfo{volume}{4}},
  \bibinfo{pages}{195} (\bibinfo{year}{2005}).

\bibitem{Dietl2002}
\bibinfo{author}{\bibfnamefont{T.}~\bibnamefont{Dietl \textit{et al.} }},
  \bibinfo{journal}{Semicond. Sci. Technol.} \textbf{\bibinfo{volume}{17}},
  \bibinfo{pages}{377} (\bibinfo{year}{2002}).

\bibitem{Ohno1998}
\bibinfo{author}{\bibfnamefont{H.}~\bibnamefont{Ohno \textit{et al.} }},
  \bibinfo{journal}{Science} \textbf{\bibinfo{volume}{281}},
  \bibinfo{pages}{951} (\bibinfo{year}{1998}).

\bibitem{Dietl2008}
\bibinfo{author}{\bibfnamefont{T.}~\bibnamefont{Dietl \textit{et al.} }},
  \bibinfo{journal}{J. Appl. Phys.} \textbf{\bibinfo{volume}{103}},
  \bibinfo{pages}{07D111} (\bibinfo{year}{2008}).

\bibitem{Sato2005}
\bibinfo{author}{\bibfnamefont{K.}~\bibnamefont{Sato \textit{et al.} }},
  \bibinfo{journal}{Jpn. J. Appl. Phys.} \textbf{\bibinfo{volume}{44}},
  \bibinfo{pages}{L948} (\bibinfo{year}{2005}).

\bibitem{Jamet2006}
\bibinfo{author}{\bibfnamefont{M.}~\bibnamefont{Jamet \textit{et al.} }},
  \bibinfo{journal}{Nature Mater.} \textbf{\bibinfo{volume}{5}},
  \bibinfo{pages}{653} (\bibinfo{year}{2006}).

\bibitem{Bougeard2006}
\bibinfo{author}{\bibfnamefont{D.}~\bibnamefont{Bougeard, S. Ahlers, A. Trampert, N. Sircar, G. Abstreiter }},
  \bibinfo{journal}{Phys. Rev. Lett.} \textbf{\bibinfo{volume}{97}},
  \bibinfo{pages}{237202} (\bibinfo{year}{2006}).

\bibitem{Li2007}
\bibinfo{author}{\bibfnamefont{A. P.}~\bibnamefont{Li \textit{et al.} }},
  \bibinfo{journal}{Phys. Rev. B} \textbf{\bibinfo{volume}{75}},
  \bibinfo{pages}{201201(R)} (\bibinfo{year}{2007}).

\bibitem{Devillers2007}
\bibinfo{author}{\bibfnamefont{T.}~\bibnamefont{Devillers \textit{et al.} }},
  \bibinfo{journal}{Phys. Rev. B} \textbf{\bibinfo{volume}{76}},
  \bibinfo{pages}{205306} (\bibinfo{year}{2007}).

\bibitem{Gu2005}
\bibinfo{author}{\bibfnamefont{L.}~\bibnamefont{Gu \textit{et al.} }},
  \bibinfo{journal}{J. Magn. Magn. Mater.} \textbf{\bibinfo{volume}{290-291}},
  \bibinfo{pages}{1395} (\bibinfo{year}{2005}).

\bibitem{Bonanni2008}
\bibinfo{author}{\bibfnamefont{A.}~\bibnamefont{Bonanni \textit{et al.} }},
  \bibinfo{journal}{Phys. Rev. Lett.} \textbf{\bibinfo{volume}{101}},
  \bibinfo{pages}{135502} (\bibinfo{year}{2008}).

\bibitem{Kuroda2007}
\bibinfo{author}{\bibfnamefont{S.}~\bibnamefont{Kuroda \textit{et al.} }},
  \bibinfo{journal}{Nature Mater.} \textbf{\bibinfo{volume}{6}},
  \bibinfo{pages}{440} (\bibinfo{year}{2007}).

\bibitem{Fiederling1999}
\bibinfo{author}{\bibfnamefont{R.}~\bibnamefont{Fiederling \textit{et al.} }},
  \bibinfo{journal}{Nature} \textbf{\bibinfo{volume}{402}},
  \bibinfo{pages}{787} (\bibinfo{year}{1999}).

\bibitem{Leycuras1995}
\bibinfo{author}{\bibfnamefont{A.}~\bibnamefont{Leycuras \textit{et al.} }},
  \bibinfo{journal}{Appl. Phys. Lett.} \textbf{\bibinfo{volume}{66}},
  \bibinfo{pages}{1800} (\bibinfo{year}{1995}).

\bibitem{Zhu2004}
\bibinfo{author}{\bibfnamefont{W.}~\bibnamefont{Zhu \textit{et al.} }},
  \bibinfo{journal}{Phys. Rev. Lett.} \textbf{\bibinfo{volume}{93}},
  \bibinfo{pages}{126102} (\bibinfo{year}{2004}).

\bibitem{Zhu2008}
\bibinfo{author}{\bibfnamefont{W.}~\bibnamefont{Zhu, Z. Zhang, E. Kaxiras }},
  \bibinfo{journal}{Phys. Rev. Lett.} \textbf{\bibinfo{volume}{100}},
  \bibinfo{pages}{027205} (\bibinfo{year}{2008}).

\bibitem{comment2}
\bibinfo{author}{\bibfnamefont{I.-S.}~\bibnamefont{Yu \textit{et al.} }},
  \bibinfo{journal}{to be published}.

\bibitem{Schulthess2001}
\bibinfo{author}{\bibfnamefont{T. C.}~\bibnamefont{Schulthess \textit{et al.} }},
  \bibinfo{journal}{J. Appl. Phys.} \textbf{\bibinfo{volume}{89}},
  \bibinfo{pages}{7021} (\bibinfo{year}{2001}).

\bibitem{Woodbury1955}
\bibinfo{author}{\bibfnamefont{H. H.}~\bibnamefont{Woodbury \textit{et al.} }},
  \bibinfo{journal}{Phys. Rev.} \textbf{\bibinfo{volume}{100}},
  \bibinfo{pages}{659} (\bibinfo{year}{1955}).

\bibitem{Achatz2008}
\bibinfo{author}{\bibfnamefont{P.}~\bibnamefont{Achatz \textit{et al.} }},
  \bibinfo{journal}{Appl. Phys. Lett.} \textbf{\bibinfo{volume}{92}},
  \bibinfo{pages}{072103} (\bibinfo{year}{2008}).


\bibitem{Solin2000}
\bibinfo{author}{\bibfnamefont{S. A.}~\bibnamefont{Solin \textit{et al.} }},
  \bibinfo{journal}{Science} \textbf{\bibinfo{volume}{289}},
  \bibinfo{pages}{1530} (\bibinfo{year}{2000}).

\bibitem{Matsukura1998}
\bibinfo{author}{\bibfnamefont{F.}~\bibnamefont{Matsukura, H. Ohno, A. Shen, Y Sugawara }},
  \bibinfo{journal}{Phys. Rev. B} \textbf{\bibinfo{volume}{57}},
  \bibinfo{pages}{R2037} (\bibinfo{year}{1998}).

\bibitem{Park2002}
\bibinfo{author}{\bibfnamefont{Y. D.}~\bibnamefont{Park \textit{et al.} }},
  \bibinfo{journal}{Science} \textbf{\bibinfo{volume}{295}},
  \bibinfo{pages}{651} (\bibinfo{year}{2002}).

\bibitem{Altshuler1985}
{B. L. Altshuler and A. G. Aronov, in $\textit{Electron - Electron
Interactions in Disordered Systems}$, edited by A. L. Efros and M.
Pollak (North-Holland, Amsterdam, 1985), p. 1.}

\bibitem{Sze1981}
{S. M. Sze, $\textit{Physics of Semiconductors Devices}$ (Wiley, New York, 1981).}

\bibitem{Polyanskaya1981}
\bibinfo{author}{\bibfnamefont{T. A.}~\bibnamefont{Polyanskaya \textit{et al.} }},
  \bibinfo{journal}{Sov. Phys. JETP Lett.} \textbf{\bibinfo{volume}{34}},
  \bibinfo{pages}{361} (\bibinfo{year}{1981}).

\bibitem{Uren1981}
\bibinfo{author}{\bibfnamefont{M. J.}~\bibnamefont{Uren \textit{et al.} }},
  \bibinfo{journal}{J. Phys. C} \textbf{\bibinfo{volume}{14}},
  \bibinfo{pages}{L395} (\bibinfo{year}{1981}).

\bibitem{Fert2001}
\bibinfo{author}{\bibfnamefont{A.}~\bibnamefont{Fert and H. Jaffr\`{e}s }},
  \bibinfo{journal}{Phys. Rev. B} \textbf{\bibinfo{volume}{64}},
  \bibinfo{pages}{184420} (\bibinfo{year}{2001}).

\end{thebibliography}
\end{document}